\documentclass[twocolumn,showpacs,preprintnumbers,pra,amsmath,amssymb]{revtex4}
\usepackage{graphicx}
\usepackage{dcolumn}
\usepackage{bm}

\def\tr{\mbox{tr}}

\def\bea{\begin{eqnarray}}
\def\eea{\end{eqnarray}}

\begin{document}

\title{Entanglement, detection, and geometry of non-classical states}

\author{Kisik Kim$^{1,2}$, Jaewan Kim$^{2}$, and Joonwoo Bae$^{2}$\footnote{bae.joonwoo@gmail.com}}

\affiliation{$^{1}$Department of Physics, Inha University, Incheon 402-751, Korea \\
$^{2}$School of Computational Sciences, Korea Institute for
Advanced Study, Seoul 130-722, Korea}

\date{\today}


\begin{abstract}
Non-classical states that are characterized by their non-positive quasi-probabilities in phase space are known to be the basis for various quantum effects. In this work, we investigate the interrelation between the non-classicality and entanglement, and then characterize the non-classicality that precisely corresponds to entanglement. The results naturally follow from two findings: one is the general structure among non-classical, entangled, separable, and classical states over Hermitian operators, and the other a general scheme to detect non-classical states.
\end{abstract}



\maketitle

\section{Introduction}

In the seminal paper of the quantum theory of light, R. Glauber has shown that quantum systems reveal their \emph{non-classicality} by non-positive quasi-probability distributions in phase space that classical systems fail to describe \cite{gs}. Not relying only on the correlational effects among quantum systems, the non-classicality was turned out to be the basis for various quantum effects. In last decades, quantum states having correlations that cannot be prepared by local operations and classical communications, i.e. entanglement, have been extensively investigated with their essential role to outperform classical counterparts in information processing.

The presence of the non-classicality is more primitive than entanglement as the non-classicality has to necessarily exist if entangled states can be generated \cite{mskim}. Or equivalently, multimode classical states can \emph{never} be entangled, and all entangled states are already non-classical. For instance, the non-classicality of the initial system is a quantity preserving under transformations via linear optical elements \cite{kim} that are often used in entanglement engineering such as quantum computation \cite{klm}. Consequently, the pre-existing non-classicality dictates or already limit the entanglement that can be manipulated. Therefore, to have a precise estimate of the non-classicality in the connection to entanglement is, not only of theoretical interest, a line that allows to decide the intrinsic capability of given quantum systems in information processing.

In this work, we characterize entanglement in terms of the non-classicality with the non-classicality measure in Ref. \cite{ctlee}. The non-classicality that precisely corresponds to entanglement is refined. The result is derived from the geometry of quantum states over quasi-probability distributions, which is based on the non-classical states detection method that we shall show later. These findings are also of fundamental importance, devoted to discovering the convex geometry of physical, separable and classical states out of Hermitian operators.

This paper is organized as follows. We first introduce a map that detects \emph{all} non-classical states. The map can also be translated to witness operators, that may be called as non-classicality witnesses. On using the map, we identify positive operators (i.e. physical states) out of positive $s$-ordered quasi-probability distributions. Remarkably, all the positive $s$-ordered quasi-probability distributions, except the normally ordered one, do not necessarily correspond to physical states. This then leads the geometric refinement of the non-classicality that finally defines the entanglement parameter: the non-classicality depth of entanglement.

\section{Detecting non-classical states}

Let us begin by introducing non-classical states and the non-classicality measure. Quasi-probability distributions in phase space are in general parameterized by the ordering parameter $s$ which takes values in $[0,1]$ and reads from the Glauber-Sudarshan $P$ function ($s=0$) to the Husimi $Q$ function (s=1). For the symmetric ordering $s=1/2$, the quasi-probability distribution corresponds to the Wigner function. For a given quantum state, when its quasi-probability distribution for any $s$ has negative values, the state is referred to as non-classical. For instance, negativity in the Wigner function has been often used as the signature of the non-classicality \cite{x}. Then in general, it holds that if a quantum state has negative probabilities in some $s$-ordered representation, its $P$ function must have negative probabilities. This can be easily seen by the regularization processing of non-positive $P$ functions, which is shown in what follows.

A multi-mode state $\rho$ can be uniquely written in terms of $P$ function as \cite{mm}, \bea \rho = \int\prod_{i=1}^{n} d^{2}{z_{i}}  P(z_{1},\cdots,z_{n}) \bigotimes_{i=1}^{n} |z_{i} \rangle\langle z_{i}|, \label{1} \eea where $|z_{i}\rangle$ in the $i$-th mode are coherent states. When the $P$ function has negative values, the state is referred to as non-classical. A non-positive $P$ function can be transformed into a true(i.e. non-negative) probability distribution by the following regularization processing: \begin{widetext}\bea R_{\tau} [P](\alpha_{1},\cdots,\alpha_{n}) = \int  \prod_{i=1}^{n} \frac{d^{2}\alpha^{'}_{i}}{\pi \tau} e^{-|\alpha-\alpha^{'}_{i}|^{2} / \tau} P(\alpha_{1},\cdots,\alpha_{n}), \label{rf}\eea\end{widetext} where $0< \tau  \leq 1$ and $R_{0}[P](\alpha) = P(\alpha)$. Note that once $R_{\tau}[P] \geq 0$ it holds that $R_{\tau^{'}}[P] \geq 0$ for all $\tau^{'}\geq \tau$, from which the minimum $\tau$ that regularizes a given $P$ function, denoted by $\tau_{m}[\rho]$ throughout the paper, has been defined as a measure to quantify the non-classicality, and called as the non-classicality depth \cite{ctleen, ctlee}. In fact, the regularization with $s$ corresponds to the transformation of quasi-probability distribution from the normally ordered representation to $s$-ordered one. The $Q$ function is given when $s = 1$, and can be expressed as, $Q(\alpha) = \langle\alpha |\rho|\alpha \rangle \geq 0$ with coherent states basis $|\alpha\rangle$. It is clear that $Q$ functions are always non-negative. Therefore, if a given Hermitian operator $\rho$ cannot be regularized with $s\leq 1$, then one can conclude that the operator is non-positive. This also means, due to the fact that the $Q$ function is non-positive, that there exists some coherent state $|\beta\rangle$ to detect a negative expectation of the $Q$ function, i.e. $Q(\beta) = \langle \beta | \rho | \beta \rangle < 0 $.


\textbf{\emph{Lemma 1.}} A Hermitian operator of unit trace is non-positive if it cannot be regularized by the transformation in Eq. (\ref{rf}).

It is clear that classical states form a convex set, since a convex combination of non-negative $P$ functions is automatically non-negative and thus constitutes a new $P$ function of the corresponding mixed state. The convex structure implies that the characterization of the set of classical states can be hugely simplified by the so-called witness operators developed much in the entanglement theory. For Hermitian operators, the Hahn-Banach theorem can be applied so that one can always find a Hermitian operator $W$ such that for all classical states $\sigma$, $\tr[W \sigma] \geq 0$ while  $\tr[W \rho] < 0$ for some non-classical ones $\rho$, which may be therefore called as non-classicality witnesses \cite{exp}.

Having collected two facts in the above, we now introduce the map that detects all non-classical states. Here, detection means that non-classical states are mapped to non-positive operators, so that the non-classical states are detected by negative expectation values. The map can be defined on the $P$ function as follows, \bea \Lambda_{a}[P(z_{i})] = P_{a}(z_{i}) = \frac{1}{a^{2}} P(\frac{z_{i}}{a}),\label{pmap} \eea which in fact describes, for $a\in[0,1]$, the state that has transmitted the beam splitter with transmittance $T=a^{2}$. The expression of the map can also be obtained on the level of states by considering state transformation under the beam splitter and then taking $a>1$. Note that non-negative $P$ functions remain non-negative under the map $\Lambda_{a}$.

\textbf{\emph{Proposition 1.}} For a non-classical state $\rho$, there exists $a > 1$ such that $\Lambda_{a}[\rho] \ngeq 0$.

\emph{Proof.} For simplicity, let us introduce the characteristic function of $\rho$ through the Fourier transformation of the $P$ function, $ \chi  (x) = \int d^{2}\alpha P(\alpha) e^{\alpha x^{*} - \alpha^{*}x} $. By regularizing $P$ function with $\tau$, the characteristic function is transformed to $K_{\tau}(\beta) = e^{-\tau|\beta|^{2}} \chi(\beta)$. The Bochner's theorem tells that the regularized function is positive if and only if the characteristic function is positive definite \cite{boc}.

Suppose that the $P$ function of a non-classical state $\rho$ can be regularized to a positive distribution with $\tau_{m}$. The map $\Lambda_{a}$ is applied to the state, and then the characteristic function denoted by $\chi_{a}(\beta)$ can be expressed as $\chi_{a}(\beta) = \chi(a \beta)$. Now, we want to see the $\tau_{a,m}$ that regularizes $\Lambda_{a}[\rho]$ on the level of characteristic function, $K_{a, \tau} (\beta) = e^{-\tau_{a}|\beta|^{2}}\chi_{a}(\beta)$. Since $K_{\tau} (\beta)$ is positive iff $\tau \geq \tau_{m}$, $K_{a, \tau} (\beta)$ is positive iff $\tau_{a} \geq a^{2} \tau_{m}$, so is $R_{\tau}[\Lambda_{a}(P)]$,  from which $\tau_{a,m} = a^{2}\tau_{m}$. If $a$ is large enough that $\tau_{a,m}>1$, this implies that $\Lambda_{a}[\rho]$ cannot be regularized and by the Lemma 1 means that non-positive. In particular, a map $\Lambda_{a}$ with $a > 1/\sqrt{\tau_{m}}$ (which is larger than $1$ since $\tau_{m}\leq 1$) can detect non-classical states having $\tau_{m}$. Since $\tau_{m} > 0$ for all non-classical states, the map $\Lambda_{a}$ can detect all non-classical states.   $\Box$

\begin{figure}
  \includegraphics[width=5.2cm]{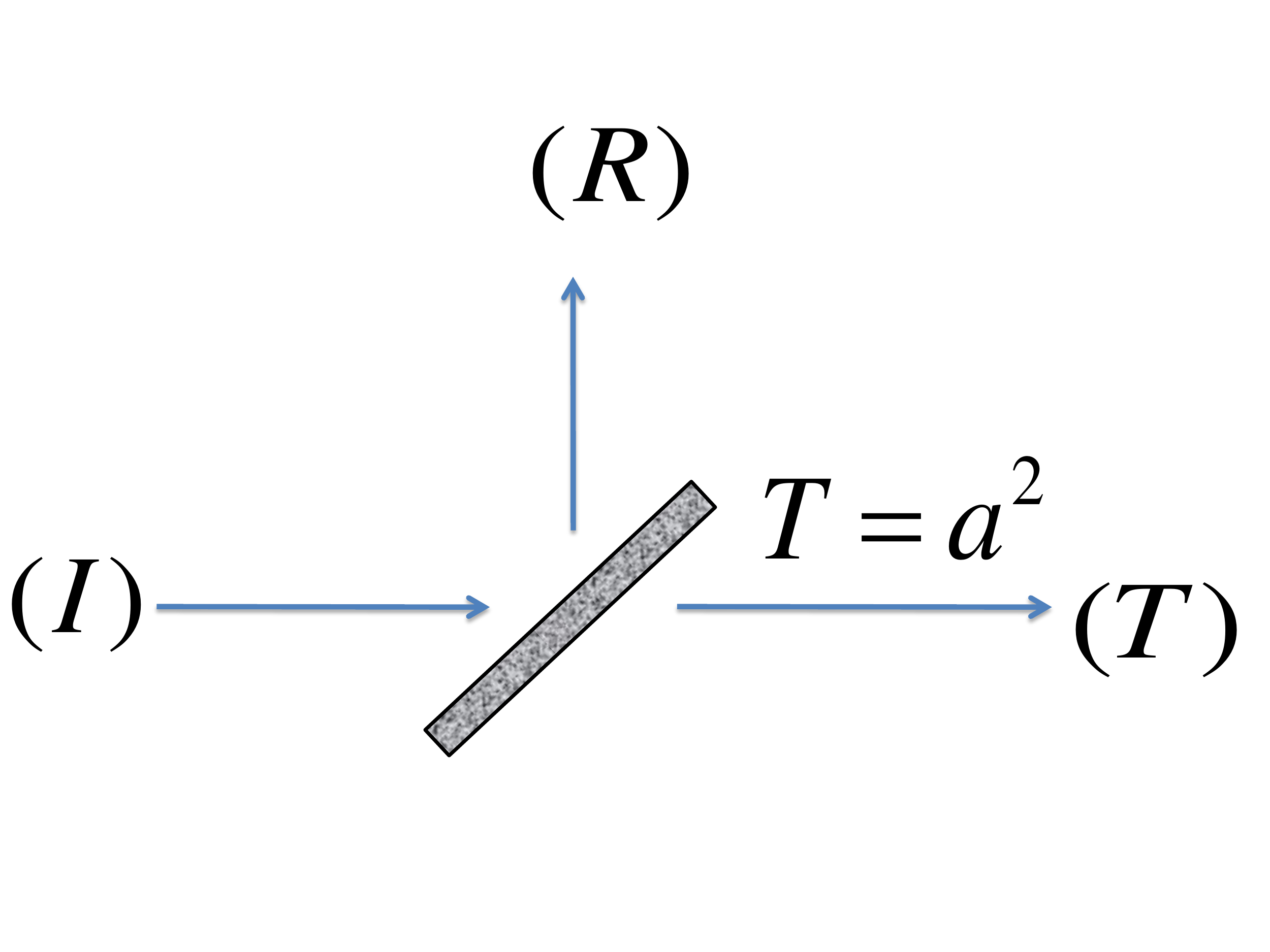}\\
   \caption{The (I), (T), and (R) are input, transmitted, and reflected states. The map in Eq. (\ref{pmap}) when $0\leq a \leq 1$ describes the relation between the input and the transmitted states. For $a>1$, the map in Eq. (\ref{pmap}) can be thought of as a "non-physical" direction from (T) to (I).}\label{beam}
\end{figure}

The proof in the above can also derive the useful relation, $\Lambda_{a}\circ R_{\tau}[P] = R_{a^{2}\tau} \circ \Lambda_{a}[P]$, that provides a geometrical structure of $s$-ordering representation as follows.

\textbf{\emph{Lemma 2.}} Let $\tau_{m}[\rho]$ denote the non-classicality depth of state $\rho$. By the map $\Lambda_{a}$, the non-classicality depth is mapped to $\tau_{m}[\Lambda_{a}[\rho]] = a^{2}\tau_{m}[\rho]$.

The idea behind the map comes up with the fact that non-physical operations may cause certain effects that cannot be interpreted being physical. For instance, in the entanglement theory, non-physical operations such as positive but not completely positive maps detect all entangled states, exploiting negative expectation values. As such, the map $\Lambda_{a}$ with $a>1$ transforms non-classical states into operators providing negative expectation values. Together with the Lemma 2, it can also be seen that the map is non-physical in that $\Lambda_{a}$ with $a>1$ describes the reverse direction from the output to the input states, increasing the non-classicality depth, which is of course non-physical as a time-reversal processing. In this way, only classical states are not detected since their non-classicality depth is constantly zero.


\textbf{\emph{Example.}} A non-classical state that is not detected by the criteria shown in Ref.\cite{vogel2} was presented in Ref.\cite{diosi}, and its $P$ function is, $P(\alpha)=\frac{2}{\pi} e^{-|\alpha|^{2}} - \delta(\alpha)$. We now apply the map $\Lambda_{a}$ to detect that $\rho$ is non-classical, \bea \langle \beta | \Lambda_{a}[\rho] | \beta\rangle = \frac{2}{a^{2}+1} \exp [ - \frac{|\beta|^{2}}{1+a^{2}}] - \exp[-|\beta|^{2}] \nonumber\eea which is non-positive for a sufficiently large $a$.$\Box$


The map in Eq. (\ref{pmap}) can be translated into, what we may call, non-classicality witnesses as follows. For $\Lambda_{a}[\rho]$ that cannot be regularized, the $Q$ function is not positive, meaning that there exists coherent state $|\beta\rangle$ such that $ Q(\beta) = \langle\beta | \Lambda_{a}[\rho] |\beta\rangle < 0$. Being constrained to keep the expectation value the same, the dual map $\Lambda_{a}^{*}$ can be obtained and applied to evolution of the coherent state, $W_{\beta} = \Lambda_{a}^{*}[|\beta\rangle\langle\beta|]$, such that the following holds \bea \tr[\rho W_{\beta}] = \tr[|\beta\rangle\langle\beta|\Lambda_{a}[\rho]].\nonumber\eea Note that the collection of all $W_{\beta}$ can completely characterize the set of classical states, since i) classical states form a convex set, and ii) $\Lambda_{a}$ detects all nonclassical states. Although coherent states are applied here due to the $Q$ function, in general, any Hermitian operators that overlap with negative ranges of $\Lambda_{a}[\rho]$ can be in the case.

\section{Geometry of non-classical states}

\begin{figure}
  \includegraphics[width=8cm]{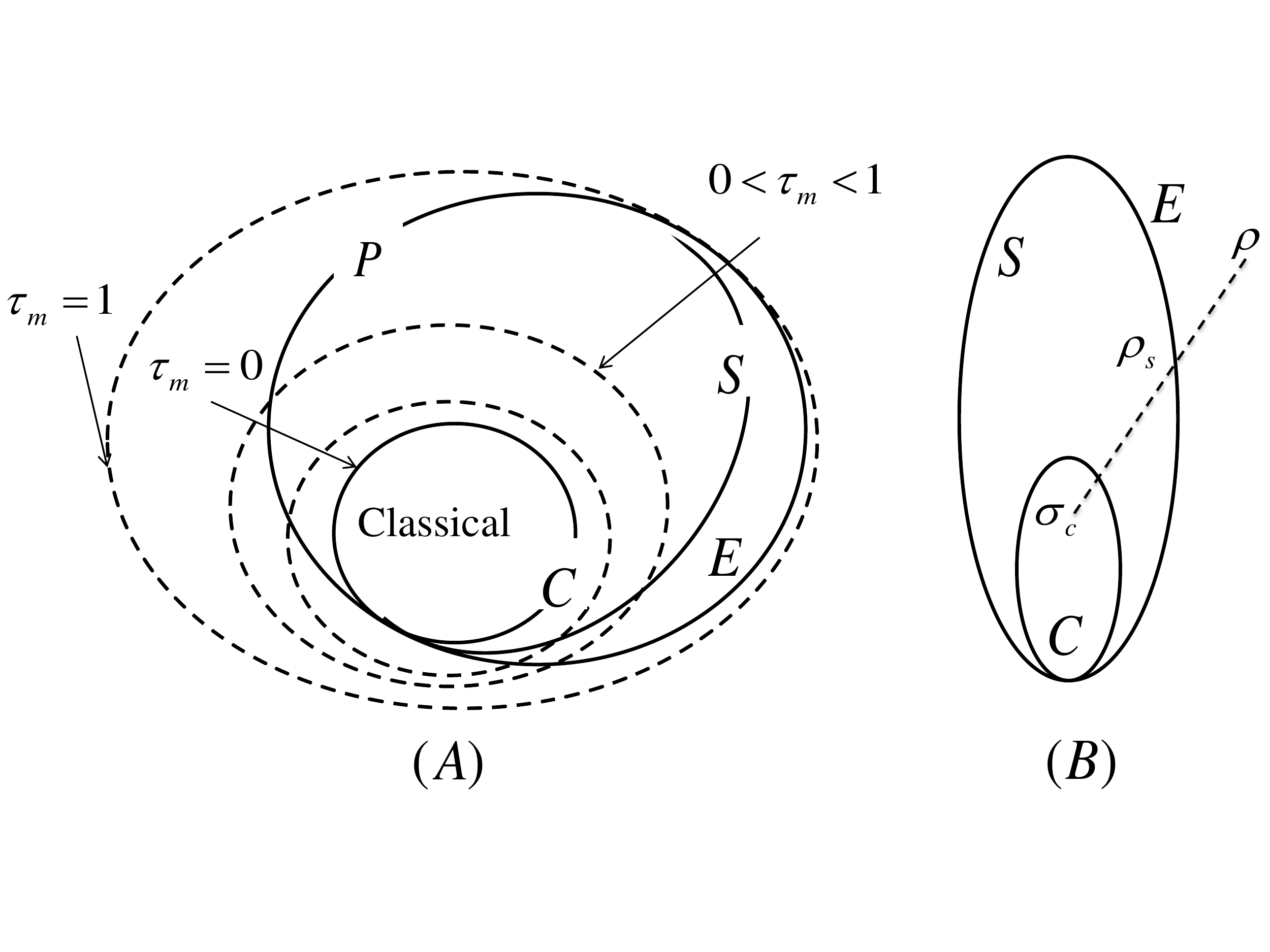}\\
  \caption{(A): Physical states are shown over the axis of the non-classicality depth. Positive operators($P$) are only a subset from $\tau=0$ to $\tau=1$, including the set of separable states($S$) which consists of all classical states. (B): The convex geometry of separable and classical states leads the entanglement parameter that takes into account the non-classicality, the NcDE in Eq. (\ref{ncde}).}\label{2clncl}
\end{figure}

So far, we have seen quantum states in terms of positive distributions in $s$-ordering representation, based on which the map $\Lambda_{a}$ with $a>1$ is shown to increase the non-classicality depth so that physical states are sent away to non-positive ones. However, positive quasi-probability distributions do not mean physical operators in general \cite{non1, non2, non3}. For instance, the $Q$ function of the following operator, $A = k|0\rangle\langle2| + |1\rangle\langle1| + k^{*}|2\rangle\langle0|$, which is the unit trace and Hermitian but not positive, has the positive $Q$ function, $Q_{A}(\beta) = |k|^{2}(\beta + \beta^{*})^{2}\geq0$. In which value of $s$ do positive quasi-probability distributions mean positive operators? In the below, $s=0$ is shown to be only the case.

First, positive operators form a convex set including separable and classical states. Let us then show that there exists a non-positive operator that can be regularized with very small $\tau$ in Eq. (\ref{rf}). The non-classical state, $\rho = (1-\epsilon)|0\rangle\langle 0| + \epsilon|2\rangle\langle 2|$, is arbitrarily close to the vacuum as $\epsilon$ tends to $0$. Applying the map, $\Lambda_{a}[\rho] = \sum_{i=0}^{2}r_{i}|i\rangle\langle i|$ with $r_{0} = (1-2\epsilon a^{2}+\epsilon a^{4})$, $r_{1} = 2\epsilon a^{2}(1-a^{2})$, and $r_{2} = \epsilon a^{4}$. Note that the coefficients $r_{i}$ are the eigenvalues of $\Lambda_{a}[\rho]$, meaning that $\Lambda_{a}[\rho]$ becomes non-positive whenever $a>1$ (since $r_{1}<0$). The regularization of $\Lambda_{a}[\rho]$ is \bea R_{\tau}[P](z) & = & \frac{1}{\tau} \exp[-\frac{|z|^{2}}{\tau}] \big( r_{0} + r_{1} (\frac{|z|^{2}}{\tau^{2}}-\frac{1-\tau}{\tau}) \nonumber \\ & + & r_{2} (\frac{|z|^{4}}{2\tau^{4}} -\frac{2(1-\tau)}{\tau^{3}}|z|^{2} +(\frac{1-\tau}{\tau})^{2}) \big). \nonumber \eea From the above, the depth of the non-classicality is \bea \tau_{m}[\Lambda_{a}[\rho]] =\frac{a^{2}\sqrt{\epsilon} }{\sqrt{1-\epsilon} + \sqrt{\epsilon}}, \nonumber \eea which can be arbitrarily small for $a>1$ by taking $\epsilon$ tends to a very small number. In summary, this shows that outside the set of positive states there exists a non-positive operator which can still have a very small non-classicality depth. This leads the following conclusion.

\textbf{\emph{Proposition 2. }} For all $s\in (0,1]$, there exist positive quasi-probability distributions that may correspond to non-positive Hermitian operators.

Based on the proposition in the above, plus that separable and classical states form convex sets, respectively, the positive operators can be drawn over the axis of the non-classicality depth as it is shown in Fig.\ref{2clncl}. Note that in general quantum states of the same non-classicality depth form a convex set. The Fig.\ref{2clncl} is drawn as well based on the following facts. First, there are separable states having the unit non-classicality depth as $\tau_{m}[\rho_{A}\otimes\rho_{B}] =1$ if and only if either $\tau_{m}[\rho_{A}] =1$ or $\tau_{m}[\rho_{B}] =1$. Separable states also consist of all classical states since classical states are already separable. Next, there are entangled states having a non-unit non-classicality depth, which can be seen by the state, \bea \rho_{p} = p |\phi^{+} \rangle \langle \phi^{+}|  + (1-p) \frac{I}{4}, \label{rhop}\eea where $|\phi^{+}\rangle = (|00\rangle + |11\rangle)/\sqrt{2}$ and $I = \sum_{i,j=0}^{1}|ij\rangle\langle ij|$. The regularization is given as \bea R_{\tau}[P](z_{A},z_{B}) & = & \frac{1}{4\tau^{2}}\exp[-\frac{|z_{A}|^{2}+|z_{B}|^{2}}{\tau}] \mathcal{A}_{\tau}(z_{A},z_{B}), \nonumber \\ \mathcal{A}_{\tau}(z_{A},z_{B}) & = & 2p\big|1+\frac{z_{A}z_{B}}{\tau^{2}}\big|^{2} + (1-p) \frac{|z_{A}z_{B}|^{2}}{\tau^{4}} \nonumber \\ && + 2p(\frac{1-\tau}{\tau})^{2} +(1-p)(\frac{2\tau-1}{\tau})^{2} \nonumber \\ &&+\frac{|z_{A}|^{2} +|z_{B}|^{2}}{\tau^{3}} (2\tau-1-p), \label{n}\eea  from which $\tau_{m}[\rho_{p}] = (1+p)/2$ for the state. it is known that the state $\rho_{p}$ is separable iff $p\leq 1/3$ \cite{ppt}. Hence, as the value $p$ decreases from $1$ to $0$, $\tau_{m}[\rho_{p}]$ does from $1$ to $1/2$, at which the state $\tau_{m}[\rho_{p}]$ passes through the border between the separable and the entangled states when $\tau=2/3$(or, equivalently, $p=1/3$). All this constitutes the geometry of physical, entangled, separable, and classical states, show in Fig.\ref{2clncl}.


\section{Entanglement of non-classical states}.

Based on the convex geometry shown in Fig.\ref{2clncl}, we are now ready to geometrically characterize what of the non-classicality corresponds to the correlational property, entanglement. This is inspired by a geometric entanglement measure, the robustness of entanglement in Ref. \cite{robu} that was based on the convexity of separable states. As such, here we are based on the convexity of classical states. For a multi-mode non-classical state $\rho$, since classical states are a subset of separable ones there always exists a separable state $\rho_{s}$ by admixing $\rho$ with a classical $\sigma$, as follows, \bea \rho_{s} = \min_{\kappa} \frac{1}{\kappa+1} ( \rho + \kappa \sigma). \label{betas}\eea Note that $\rho_{s}$ lies on the boundary of separable states in Fig.\ref{2clncl}. Using the state $\rho_{s}$, one can divide the non-classicality depth into two: one from entangled to separable states, and the other from the separable to classical ones.

\emph{\textbf{Definition.}} [The non-classicality depth of entanglement (NcDE)] The NcDE of $\rho$ with respect to the classical state $\sigma$, denoted by $N_{e}(\rho\| \sigma)$, is $ N_{e}(\rho\| \sigma) = \tau_{m}[\rho]- \tau_{m}[\rho_{s}]$, with the state $\rho_{s}$ found by Eq. (\ref{betas}). The NcDE of a state $\rho$ is, \bea N_{e}(\rho) = \min_{\sigma\in C} N_{e}(\rho\| \sigma), \label{ncde}\eea where the minimization runs over the set of all classical states $C$.


It is clear that by definition the NcDE of separable states is zero. Then, for an entangled state $\rho$, the non-classical depth is strictly larger than the $\rho_{s}$ defined in Eq. (\ref{betas}) for all classical states $\sigma$. This can be seen by the inequality that for non-classical $\rho$ and classical $\sigma$, it holds $\tau_{m}[\rho] > \tau_{m} [(1-\epsilon)\rho + \epsilon \sigma]$ for all $0< \epsilon \leq 1$. The details are shown in the appendix.  Hence, the NcDE is an entanglement parameter firstly derived from non-classicality of quantum states.

In general, to obtain the NcDE, one should minimize the $N(\rho\|\sigma)$ over all classical states, and then has to apply a separability criteria to obtain a separable state $\rho_{s}$ that is interpolated by $\rho$ and $\sigma$. Moreover, the optimization processing in the NcDE runs for all bosonic systems including non-Gaussian states, for which little is known about the separability criteria. It is thus generally hard to explicitly evaluate.

\begin{figure}
  \includegraphics[width=8cm]{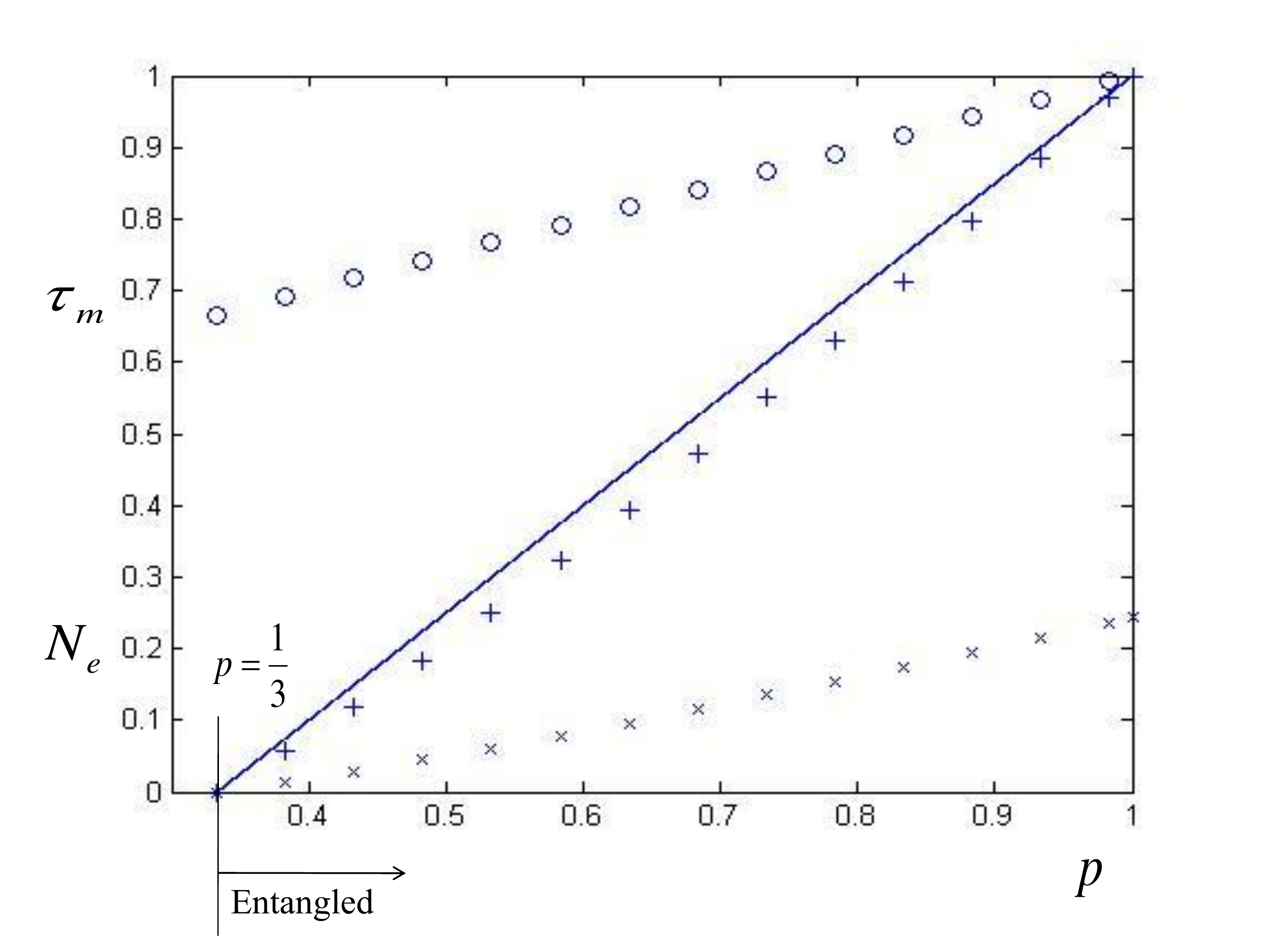}\\
  \caption{The NcDE (plotted by $\times$) $N_{e}$of the state $\rho_{p}$ is fitted. Note that $\rho_{p}$ is entangled iff $p \geq 1/3$. The circle shows the non-classicality depth $\tau_{m}$ of the state for each $p$: $\tau_{m}[\rho_{p}] = (1+p)/2$. The normalized NcDE is plotted by $+$, which is proportional to the entanglement parameter, negativity (solid line), $\mathcal{N}(\rho_{p}) = (3p-1)/2$. }\label{plot}
\end{figure}

To illustrate the NcDE with an example, let us now explicitly compute it for the state $\rho_{p}$ in Eq.(\ref{rhop}). Since the optimization is hard in general, the NcDE is computed here with respect to the ansatz state, $\sigma$ as follows. As it can be seen in Eq. (\ref{n}), the state $\sigma$ that classicalizes $\rho_{p}$ should remove the non-positive part of the $P$ function of $\rho_{p}$. That is, the $P$ function of $\sigma$ is \bea P_{\sigma}(z_{A},z_{B}) = \frac{|z_{A}|^{2} + |z_{B}|^{2}}{2\pi^{2}\tau^{3}} \exp[-\frac{|z_{A}|^{2} + |z_{B}|^{2}}{\tau}].  \label{psig}\eea Let $\rho(\beta)$ denote the mixture of $\rho_{p}$ with the classical state $\sigma$, $\rho(\beta) = (\rho_{p} + \beta \sigma)/(1+\beta)$, which is classical if and only if $ \beta \geq \pi^{2}\tau^{-2}(1+p-2\tau)/2 $. Then, one can find the minimal value of $\beta_{s}$ such that $\rho(\beta_{s})$ is separable. Note that $\rho_{p}$ is defined in $2\otimes 2$ dimensional Hilbert space and $\sigma$ is separable. Therefore, decomposing the state $\sigma$ with two-dimensional number basis $\{|0\rangle,|1\rangle\}$ and the rest, i.e. $(2\otimes 2)$ subsystems are only relevant, one can apply the known separability criteria \cite{ppt}. Finally,  \bea \beta_{s} = \frac{(3p-1)(1+\tau)^{4}}{6\tau}. \label{minbs}\eea The non-classicality depth of the state, $\tau_{m}[\rho(\beta_{s})]$, can be obtained by numerics, and the NcDE of the state is plotted in Fig.\ref{plot}. The NcDE behaves similarly with the known entanglement measure, the negativity.

\section{Conclusion}

To conclude, we provide the general method of detecting non-classical states and find the geometry of physical states over positive $s$-ordered quasi-probability distributions. It is shown that positive $s$-ordered quasi-probability distribution can generally correspond to non-positive operators. Together with the convexity of positive operators, the set of positive (i.e. physical) states are characterized. Based on the geometry, we have finally derived the entanglement parameter from the non-classicality, the NcDE.

\section*{Acknowledgement}

We thank W. Vogel and M. Piani for helpful comments and discussions. This work is supported by the Korea Research Foundation Grant funded by the Korean Government (MOEHRD, Basic Research Promotion Fund) under the contract number: KRF-2009-0070-885 and KRF-2008-313-C00185.

\section*{Appendix}

We prove the inequality for a non-classical state $\rho$ and a classical one $\sigma$, \bea  \tau_{m}[\rho] > \tau_{m}[ (1 - \epsilon)\rho + \epsilon\sigma] \label{cla} \eea for $0< \epsilon \leq 1$, where $\tau_{m}$ defined as the minimum amount of thermal noise in the regularization processing: \bea R_{\tau} [P] (\alpha) = \int \frac{d^{2}\alpha'}{\pi\tau} e^{\frac{|\alpha - \alpha'|^{2}}{\tau}} P(\alpha'). \label{regularization} \eea

First, let $P_{\rho}(z)$ denote the $P$-function of a non-classical state $\rho$, for which there exists the minimum value $\tau_{1}$ that the $P$-function is regularized, i.e. $R_{\tau_{1}}[P_{\rho}](z) \geq 0$. Hence, we have $\tau_{1} = \tau_{m}[\rho]$.

Let $\tau_{2}$ denote $\tau_{m}[(1-\epsilon) \rho + \epsilon \sigma]$ for $0 < \epsilon \leq 1$ and a classical state $\sigma$, i.e. $R_{\tau_{2}}[(1-\epsilon) P_{\rho} + \epsilon P_{\sigma}] \geq 0$, where $P_{\sigma}$ is the $P$-function of $\sigma$. Note that since $\sigma$ is classical $R_{\tau}[\sigma]$ is positive for all $\tau \geq 0$. Also note that the regularization is linear, $R_{\tau_{2}}[ (1-\epsilon) P_{\rho} + \epsilon P_{\sigma}] = (1-\epsilon) R_{\tau_{2}}[P_{\rho}] + \epsilon R_{\tau_{2}}[P_{\sigma}]$. This means that, by $\tau_{2}$ in (\ref{regularization}), the function $P_{\rho}$ is not yet regularized but transformed such that \bea  R_{\tau_{2}}[P_{\rho}] \geq - \frac{\epsilon}{1-\epsilon} R_{\tau_{2}} [P_{\sigma}]. \label{eq1} \eea The rhs in (\ref{eq1}) is negative, meaning again that by $\tau_{2}$ in (\ref{regularization}) the $P$-function $P_{\rho}$ is not regularized. Therefore, there exists the minimum value $\tau > 0$ that regularizes $R_{\tau_{2}}[P_{\rho}]$, i.e. \bea R_{\tau}[R_{\tau_{2}}[P_{\rho}]] \geq 0. \eea Let us recall the identity $R_{a+b} [P] = R_{a}[R_{b} [P]]$. Hence, we arrive at the identity, for $\tau>0$, $\tau_{1} = \tau + \tau_{2}$. It is thus proved that $\tau_{m}[\rho] > \tau_{m}[(1-\epsilon) \rho + \epsilon\sigma]$.


\begin{thebibliography}{99}
\bibitem{ctleen}
This is a generalization of the non-classicality depth of single-mode states in Ref. \cite{ctlee}, for multi-mode states.
\bibitem{exp}
Note that the witnesses proposed in \cite{jk} are non-linear, and here witnesses derived by $\Lambda^{*}$ are \emph{linear} since $\Lambda_{a}$ in Eq. (\ref{pmap}) is linear.
\end{thebibliography}
\end{document}